\documentclass[a4paper,12pt]{article}
\usepackage{amsmath}
\usepackage{graphicx,psfrag,epsf}
\usepackage{subcaption}
\usepackage{enumerate}
\usepackage{natbib}
\usepackage{setspace}
\usepackage{anysize}
\usepackage{amsthm}
\usepackage{multirow}
\usepackage{amssymb}
\usepackage{arydshln}

\newcommand{\blind}{0}

\renewcommand\footnotemark{}
\newcommand{\bC}{{\bf C}}
\newcommand{\bD}{{\bf D}}
\newcommand{\bG}{{\bf G}}
\newcommand{\bH}{{\bf H}}
\newcommand{\bI}{{\bf I}}

\newcommand{\bM}{{\bf M}}
\newcommand{\bP}{{\bf P}}
\newcommand{\bQ}{{\bf Q}}
\newcommand{\bR}{{\bf R}}
\newcommand{\bS}{{\bf S}}

\newcommand{\bU}{{\bf U}}
\newcommand{\bV}{{\bf V}}
\newcommand{\bW}{{\bf W}}
\newcommand{\bX}{{\bf X}}
\newcommand{\bY}{{\bf Y}}
\newcommand{\bZ}{{\bf Z}}
\newcommand{\bff}{{\bf f}}
\newcommand{\bg}{{\bf g}}
\newcommand{\bt}{{\bf t}}
\newcommand{\bw}{{\bf w}}
\newcommand{\bo}{\boldsymbol}

\marginsize{1.6 cm}{1.6cm}{1.9 cm}{3.7 cm}
\begin{document}

\if0\blind
{
  \title{\bf Robust split-plot designs for model misspecification%\thanks{Dr. Lin is a Professor in the Department of Applied Mathematics and Institute of Statistics. He is the corresponding author of this paper. His email address is chlin6@nchu.edu.tw.}
  }
   \author{Chang-Yun Lin\\    \small \it Department of Applied Mathematics and Institute of Statistics,\\ \small \it National Chung Hsing University, Taichung, Taiwan, 40227\\
 }
    \date{} 
 \maketitle
} \fi

\begin{abstract}
%Many approaches for constructing optimal split-plot designs have been proposed. However, the optimal designs obtained by these methods may not be robust. 
%For instance, the $D$-optimal split-plot design may have higher bias for the estimation and the minimum aberration split-plot design could have higher variances and covariance for the estimation.  
%In this paper, we propose a new method to construct optimal split-plot designs that are 
Many existing methods for constructing optimal split-plot designs, such as $D$-optimal designs, only focus on minimizing the variances and covariances of the estimation for the fitted model. 
However, the underlying true model is usually complicated and unknown and the fitted model is often misspecified. 
If there exist significant effects that are not included in the model, then the estimation could be highly biased.
Therefore, a good split-plot designs should be able to simultaneously control the variances/covariances and the bias of the estimation. 
%Robust split-plot designs are rarely discussed in literature. 
%In practice, the underlying true model for an experiment is unknown and hence the fitted model is usually misspecified. 
%If there exist significant effects that are not included in the model, then the estimation of effects could be highly biased. 
In this paper, we propose a new method for constructing optimal split-plot designs that are robust for model misspecification.
%We extend the application of the $D$-optimal minimax criterion to the split-plot design and provide a general form of the loss function for the criterion.
We provide a general form of the loss function used for the $D$-optimal minimax criterion and apply it to searching for robust split-plot designs. 
To more efficiently construct designs, we develop an algorithm which combines the anneal algorithm and point-exchange algorithm. 
%We develop an algorithm which combines the anneal algorithm and point-exchange algorithm to construct the designs. 
We modify the update formulas for calculating the determinant and inverse of the updated matrix and apply them to increasing the computing speed for our developed program.   

\end{abstract}


\section{Introduction}\label{se:introduction}
%Completely randomized designs are often suggested in designs of experiments to avoid bias  the effects of possible   
%1.introduce s-p design
In experimental designs, the completely randomization is usually  recommended to avoid the bias caused by the factors that are not controlled. However, in many industrial or agricultural experiments, there often exist factors whose levels are difficult to change. These factors are called the whole-plot factors. The other factors whose levels are easy to change are called the subplot factors. If whole-plot factors exist in an experiment and the completely randomization is conducted, then the experimental cost will increase due to frequently changing levels of the whole-plot factors. To reduce the cost, the two-stage randomization strategy is usually suggested. 
First the randomization is conducted only for the treatments of whole-plot factors, call the whole plots. 
Then in each whole plot, the second randomization is conducted for the treatments of subplot factors, called the subplots. 
This kind of experiments was first introduced by Fisher (1925) and the design used for the experiment is referred to as the split-plot design (SPD).
%2.How to choose or construct the best split-plot designs has been widely discussed in literature.
%2.1. from existing design. PB, CCD, 
%2.2. another $D$-eff design (model based)
%2.3 ee design

The two-stage randomization results in the multistratum structure of the split-plot design and leads to two error terms, the whole-plot errors and the subplot errors.
%Because of the two-stage randomization, the split-plot experiment results in multistratum structure and leads to two error terms, the whole-plot errors and the subplot errors. 
Therefore, traditional analysis methods for completely randomized designs are no longer appropriate and the optimal completely randomized designs may not be optimal for the split-plot experiment. 
In literature, many approaches for constructing optimal split-plot designs have been proposed. 
A common construction method is based on the minimum aberration criterion, which can be found in Huang et al. (1998), Binham and Sitter (1999a, 1999b, 2001, 2003), Mukerjee and Fang (2002), and Tichon et al. (2012). %, and Po and Lin (2015).  
The minimum aberration criterion aims to find the optimal design which minimizes the alias of the important effects. 
An idea of this method is that the powers to detect significant effects are not the same for whole-plot factors and subplot factors. It is usually assumed that the subplot variability is smaller than the whole-plot variability, which implies that the power to detect significant subplot effects is greater than the power to detect significant whole-plot effects. Hence, subplot factors are considered more important and should be given shorter world length than whole-plot factors. The design that sequentially minimizes the wordlength patterns is selected as the minimum aberration split-plot design.
Since the minimum aberration criterion minimizes the alias, the optimal split-plot design constructed by this criterion should have less bias of the estimation for the important effects.   

%Another widely used strategy is to construct the split-plot design which has higher $D$-efficiency. This strategy is based on a given model, usually the second-order model. A good design should have good estimation ability, which can be evaluated by measuring the determinant of the information matrix of factor effects. A design is more efficient

Another widely used criterion for constructing optimal split-plot designs is the $D$-optimal criterion, which can be found in Lesinger et al. (1996), Goos and Vandebroek (2001, 2003, 2004) and Jones and Goos (2007).
The basic idea of this construction method is that a good split-plot design should have higher estimation ability for the fitted model. %, i.e., it should minimize the variances and covariances of the estimation of the effects. 
This estimation ability can be evaluated by measuring the determinant of the information matrix of a design, called the $D$-efficiency. 
The $D$-optimal split-plot design is the one which has the highest $D$-efficiency among all of designs.
Since maximizing the $D$-efficiency of a design is equivalent to minimizing the determinant of the variance-covariant matrix of the estimation, the $D$-optimal split-plot design should have smaller variances and covariances for the estimation of the effects. A good review of split-plot designs can be found in Jones and Nachtsheim (2009).

%A split-plot design is said more efficient  if its $D$-efficiency is the highest among all of possible designs. Since maximizing the the determinant of its information matrix is equivalent to minimizing the covariant of  the estimation. 

%This strategy is based on a given model, usually the second-order model. A good design should have good estimation ability, which can be evaluated by measuring the determinant of the information matrix of factor effects.  A design is said more efficient on estimation if the determinant of its information matrix is higher. 
%
%Maximizing the the determinant of its information matrix is equivalent to minimizing the covariant of  the estimation.
%
%3. based on the fitted model. true model is unknown. 

Although the minimum aberration criterion and the $D$-optimal criterion are commonly used for constructing optimal split-plot designs, the optimal designs selected by the two criteria may have higher mean square error (MSE), which consists of the square of the bias matrix and the variance-covariance matrix of the estimation.  
%The two criterion introduced above have their own drawback. 
The minimum aberration criterion focuses on minimizing the bias of the estimation but less considers minimizing the variances and covariances. Therefore, the optimal split-plot design constructed by this criterion may have higher mean square error due to higher variances or covariances of the estimation. 
On the contrary, the $D$-optimal criterion aims to minimize the variances and covariances but ignores the bias of the estimation. When the fitted model is 
misspecified, there exists a bias for the estimation. Therefore, the optimal split-plot design constructed by the $D$-optimal criterion may have higher mean square error due to higher bias of the estimation.
%
%The mean square error is compose of the variance-covariance matrix and the bias matrix. 
% 
%While the fitted model is correct, there is no bias for the estimation. Therefore, maximizing the the determinant of its information matrix is equivalent to minimizing the covariant of the estimation. 
%
%However, when the fitted model is misspecified, there exist a bias for the estimation. Therefore, The mean square error is compose of the variance-covariance matrix and the bias matrix. The traditional $D$-efficiency criterion only to minimizing the variance covariance without considering the bias of the estimate. Therefor, the optimal $D$-optimal design may have higher bias and hence has higher mean square error.
%
%4. In this paper we provide misspecification. extend minimax.

In this paper, we take the model misspecification into account. When the fitted model differs from the underlying true model, a good split-plot design should be able to simultaneously control the variance-covariance matrix and the bias matrix of the estimation. 
An appropriate criterion to deal with the model misspecification is the $D$-optimal minimax criterion proposed by Zhou (2001, 2008), Wilmut and Zhou (2011), Lin and Zhou (2013) and Yin and Zhou (2014). This criterion is usually applied on the construction of the robust completely randomized design for model misspecification. In this paper, we extend the application of the $D$-optimal minimax criterion to the split-plot designs and provide a general form of the loss function used by this criterion. This general form of the loss function allows us apply the $D$-optimal minimax criterion to selecting optimal design for split-plot experiments or completely randomized experiments with or without model misspecification. 
%to be suitable for selecting optimal split-plot design. 
%6 algorithm 
To more efficiently construct and search for the robust split-plot design, we combine the point-exchange algorithm proposed by Goos and Vandebroek (2001) and the anneal algorithm proposed by Zhou (2001). The update formulas suggested by Arnouts and Goos (2010) are applied to increasing the computing speed for calculating the loss function for the $D$-optimal minimax criterion.  

The rest of this paper is organized as follows. Section~\ref{se:cri} introduces the split-plot design and the $D$-optimal minimax criterion with the general form of the loss function, which can be used for selecting robust split-plot designs or completely randomized designs. Section~\ref{se:au} provides an algorithm which combines the point-exchange algorithm and the anneal algorithm for constructing and selecting the robust split-plot design for model misspecification. The update formulas are introduced and modified for the $D$-optimal minimax criterion. Section~\ref{se:ex} provides two examples to demonstrate how to apply our proposed method to obtaining robust split-plot designs. Section~\ref{se:co} is the conclusions and remarks.

\section{Background and criterion}
\label{se:cri}
Let $\cal H$ denote an $N$-run full factorial design for factors $F_1,\cdots,F_m$ with levels $s_1,\cdots,s_m$, respectively, where $N=\prod_{i=1}^ms_i$ and the levels of factors are coded as orthogonal contrasts.   
Let $\bH$ be the $N\times N$ matrix whose first column is all ones for the grand mean and the other $N-1$ columns are the contrasts of all the main effects and interactions of the full factorial design. 
The $i$th row of $\bH$ is corresponding to the $i$th design point (run) in $\cal H$. 

\subsection{Estimation of the split-plot design}
A split-plot design with $n$ runs can be selected from the $N$ rows of $\cal H$ without replacement by arranging the design points that have the same level combinations of the whole-plot factors into a whole plot. 
Assume that $\cal D$ is an $n$-run split-plot design with $m_w$ whole-plot factors and $m_s=m-m_w$ subplot factors, where the total number of whole plots is $b$ and the number of subplots in the $i$th whole plot is $n_i$, $i=1,\cdots,b$. 
Let $R$ be a requirement set containing $p$ effects which usually includes all the main effects of whole-plot factors and subplot factors and some interactions. 
%Let the $p$ represents the number of effects in $R$. 
Then the linear model for $R$ is
\begin{equation}\label{eq:m1}
\bY=\bX_1\bo\beta_1+\bZ\bo\gamma+\bo\epsilon,
\end{equation}
where $\bY$ is the $n\times 1$ vector of response, $\bo\beta_1$ is the $(1+p)\times 1$ vector of the grand mean and the effects in $R$, $\bX_1$ is the $n\times (1+p)$ matrix of the orthogonal contrasts for $\bo\beta_1$, $\bZ$ is an $n\times b$ indicator matrix with entries $z_{li}=1$ if the $l$th run of $\cal D$ belongs to the $i$th whole-plot and $z_{li}=0$ otherwise, $\bo\gamma$ is the $b\times 1$ vector of random whole-plot errors, and $\bo\epsilon$ is the $n\times 1$ vector of random subplot errors. It is assumed that $\bo\gamma$ and $\bo\epsilon$ are independent and have mean zero and variance-covariance matrix $\sigma_\gamma^2\bI_b$ and $\sigma_\epsilon^2\bI_n$, respectively.
 
%Since there exists a multistratum structure for a split-plot design, the variance-covariance matrix of $\bY$ is
%\begin{equation}\label{eq:sigma.spd}
%{\boldsymbol\Sigma}=\sigma_\epsilon^2({\bf I}_N+d\bf{V}),
%\end{equation}
%where $d=\sigma_\gamma^2/\sigma_\epsilon^2$, and
%\begin{equation}\label{eq:v}
%{\bf V=ZZ'}=\left[\begin{array}{cccc}{\bf 1}_{n_1} {\bf 1}'_{n_1}& {\bf 0}_{n_1} {\bf 0}'_{n_2} & \cdots & {\bf 0}_{n_1} {\bf 0}'_{n_w} \\{\bf 0}_{n_2} {\bf 0}'_{n_1} & {\bf 1}_{n_2} {\bf 1}'_{n_2} & \cdots & {\bf 0}_{n_2} {\bf 0}'_{n_w} \\ \vdots & \vdots & \ddots & \vdots \\{\bf 0}_{n_w} {\bf 0}'_{n_1} & {\bf 0}_{n_w} {\bf 0}'_{n_2} & \cdots & {\bf 1}_{n_w} {\bf 1}'_{n_w}\end{array}\right].
%\end{equation}
%The ${\bf 1}_{n_i}$ and ${\bf 0}_{n_i}$ are the $n_i\times 1$ vectors of ones and zeros for $i=1,2,\cdots,w$, respectively. 

Since there exists a multistratum structure for the split-plot design, the variance-covariance matrix of $\bY$ is
\begin{equation}\label{eq:sigma.spd}
{\boldsymbol\Sigma}=\sigma_\epsilon^2({\bf I}_n+d\bf{ZZ'}),
\end{equation}
where $d=\sigma_\gamma^2/\sigma_\epsilon^2$. 
If the entries of $\bY$ are grouped per whole plots, then equation (\ref{eq:sigma.spd}) can be written as the $n\times n$ block diagonal matrix
\begin{equation*}\label{eq:v}
{\boldsymbol \Sigma}=\left[\begin{array}{cccc}\boldsymbol \Sigma_1& {\bf 0}& \cdots & {\bf 0}\\
{\bf 0}&\boldsymbol \Sigma_2 & \cdots & {\bf 0} \\
 \vdots & \vdots & \ddots & \vdots \\
 {\bf 0} & {\bf 0} & \cdots & \boldsymbol \Sigma_b
 \end{array}\right],
\end{equation*}
where 
$\boldsymbol \Sigma_i=\sigma_\epsilon^2(\bI_{n_i}+d{\bf 1}_{n_i}{\bf 1}_{n_i}')$ for $i=1,\cdots, b$.
The generalized least square estimate (GLSE) of $\bo\beta_1$ is
\begin{equation*}\label{eq:beta.spd}
\hat{\boldsymbol \beta}_1=({\bX_1'\bo\Sigma}^{-1}{\bX_1})^{-1}{\bX_1}'{\bf \Sigma}^{-1}{\bf Y},
\end{equation*}
and the variance-covariance matrix of $\hat{\bo\beta}_1$ is
\begin{equation*}\label{eq:cov.spd}
cov(\hat{\boldsymbol\beta}_1)=({\bX_1'\bo\Sigma}^{-1}{\bX_1})^{-1}.
\end{equation*}
The estimates of $\sigma_\gamma^2$ and $\sigma_\epsilon^2$ can be obtained by the restricted maximum likelihood (REML) method (see Letsinger et al., 1996). %When all of factors are easy to change, i.e., $b=0$, the split-plot designs reduce to the completely randomized designs. 

\subsection{General form of the $D$-optimal minimax criterion}\label{se:ld}
Let $\bH=(\bH_1,\bH_2)$, where $\bH_1$ is the $N\times(p+1)$ submatrix of $\bH$ with the column ${\bf1}_N$ for the grand mean and the columns for the $p$ effects in the requirement set $R$, and $\bH_2$ is the $N\times(N-p-1)$ submatrix of $\bH$ with the columns for the effects not in $R$. Since the columns of $\bH$ are orthogonal, it is obvious that 
\[
\bH'\bH=\left(\begin{array}{cc}\bV_1&\bf 0\\\bf 0&\bV_2\end{array}\right),
\]
where both $\bV_1=\bH_1'\bH_1$ and $\bV_2=\bH_2'\bH_2$ are diagonal matrices. 
If there exist significant effects that are not included in $R$, then model (\ref{eq:m1}) is misspecified. The underlying true model with small departures from (\ref{eq:m1}) can be written as
\begin{equation*}
{\bf Y=X}_1\boldsymbol\beta_1+{\bf X}_2{\boldsymbol\beta_2}+{\bf Z}\boldsymbol\gamma+\boldsymbol\epsilon,
\end{equation*}
where $\bo\beta_2$ is the $(N-p-1)\times 1$ vector of all the effects not in $R$ and $\bX_2$ is the $n\times(N-p-1)$ matrix of the orthogonal contrasts for $\bo\beta_2$. The unknown parameter vector is assumed satisfying $\frac{1}{N}\bo\beta_2'\bV_2\bo\beta_2\leq\alpha^2$, where $\alpha\geq 0$ controls the seriousness of departures. Note that model (\ref{eq:m1}) is correct if $\alpha=0$.

When model (\ref{eq:m1}) is misspecified $(\alpha>0)$, the generalized least square estimate of $\bo\beta_1$ is biased with
\[
\begin{array}{rl}
bias(\hat{\bo\beta_1})&=E(\hat{\bo\beta_1})-\bo\beta_1\\
&=({\bf X}_1'{\bf \Sigma}^{-1}{\bf X}_1)^{-1}{\bf X}_1'{\boldsymbol \Sigma}^{-1}{\bf X}_2{\boldsymbol\beta}_2.\\
\end{array}
\]
Then the mean square error of $\hat{\bo\beta}_1$ is
\begin{equation}\label{eq:mse}
\begin{array}{rl}
MSE(\hat{\bo\beta}_1,\bX_1,\bo\beta_2)&=cov(\hat{\boldsymbol\beta_1})+bias(\hat{\bo\beta_1})bias(\hat{\bo\beta_1})'\\
&=({\bf X_1'\Sigma}^{-1}{\bf X_1})^{-1}+({\bf X}_1'{\bf \Sigma}^{-1}{\bf X}_1)^{-1}{\bf X}_1'{\boldsymbol \Sigma}^{-1}{\bf X}_2{\boldsymbol\beta}_2{\boldsymbol\beta}_2'{\bf X}_2'{\boldsymbol \Sigma}^{-1}{\bf X}_1({\bf X}_1'{\bf \Sigma}^{-1}{\bf X}_1)^{-1}.\\
\end{array}
\end{equation}
A robust split-plot design should be able to simultaneously control the variances/covariances and the bias of the estimation. To construct and obtain the robust split-plot design for model misspecification, we adopt the $D$-optimal minimax criterion and provide a general form of the loss function which can be applied to searching for the optimal design with or without multistratum structure.
%modify  to be applicable for both design with or without the multistratum structure. Since the minimax criterion  proposed by xxx is a special case of our proposed version, we call the new version the generalized $D$-optimal minimax criterion. It is described as follows.
Define the loss function of design $\cal D$ with respect to the requirement set $R$ as 
\begin{equation}\label{eq:ld0}
L_R({\cal D})=\max_{\bo\beta_2\in\Theta}|MSE(\hat{\bo\beta}_1,\bX_1,\bo\beta_2)|,
\end{equation}  
where $\Theta=\{\bo\beta_2|\frac{1}{N}\bo\beta_2'\bV_2\bo\beta_2\leq\alpha^2\}$ and $|\cdot|$ is the determinant of a matrix.
Let the singular value decomposition (SVD) of $\bo\Sigma$ be $\bU\Lambda\bU'$ %where $\bU$ is an $n\times n$ unitary matrix 
where $\bU$ is an $n\times n$ unitary matrix and $\Lambda$ is an $n\times n$ diagonal matrix with non-negative eigenvalues of $\bo\Sigma$. Define $\bo\Sigma^{-2}=\bU\Lambda^{-2}\bU'$. Then
Equation (\ref{eq:ld0}) can be written as 
\begin{equation}\label{eq:ld}
L_R({\cal D})=\frac{1+N\alpha^2\lambda_{max}\left(\bV_1^\frac{1}{2}(\bX_1'{\bf \Sigma}^{-1}{\bf X}_1)^{-1}\bX_1'{\bf \Sigma}^{-2}{\bf X}_1\bV_1^{-\frac{1}{2}}-\bV_1^{-\frac{1}{2}}\bX_1'{\bf \Sigma}^{-1}\bX_1\bV_1^{-\frac{1}{2}}\right)}{|\bX_1'{\bf \Sigma}^{-1}{\bf X}_1|},
\end{equation}
where $\lambda_{max}(\cdot)$ is the greatest eigenvalue of a matrix.
The split-plot design that minimizes the loss function among all of the possible designs is call the $D$-optimal minimax split-plot designs, which is robust for model misspecification.

Equation (\ref{eq:ld}) is a general form for the $D$-optimal minimax criterion which can be applied on both split-plot designs and completely randomized designs with or without model misspecification as follows. 
\begin{itemize}
\item[I.] When $d=0$, $\bo\Sigma$ reduces to $\sigma_\epsilon^2\bI_n$ and equation (\ref{eq:ld}) reduces to 
\[
\begin{array}{rl}
L_R({\cal D})&=\frac{1+N\alpha^2\lambda_{max}(\sigma_\epsilon^{-2}\bI_n-\sigma_\epsilon^{-2}\bV_1^{-\frac{1}{2}}\bX_1'\bX_1\bV_1^{-\frac{1}{2}})}{\sigma_{\epsilon}^{-2(p+1)}|\bX_1'\bX_1|}\\
&=\sigma_{\epsilon}^{2(p+1)}\frac{1+\frac{N\alpha^2}{\sigma_\epsilon^{2}}\left\{1-\lambda_{min}(\bV_1^{-\frac{1}{2}}\bX_1'\bX_1\bV_1^{-\frac{1}{2}})\right\}}{|\bX_1'\bX_1|},\\
\end{array}
\]
where $\lambda_{min}(\cdot)$ is the smallest eigenvalue of a matrix. Therefore, equation (\ref{eq:ld}) reduces to the form of the $D$-optimal minimax criterion given in Lin and Zhou (2013) for selecting the robust completely randomized design with model misspecification.

\item[II.] When $\alpha=0$, equation (\ref{eq:ld}) reduces to $L_R({\cal D})=1/|\bX_1'\bo\Sigma^{-1}\bX_1|$. Minimizing $L_R({\cal D})$ is equivalent to maximizing $|\bX_1'\bo\Sigma^{-1}\bX_1|^{1/(p+1)}$ and hence the $D$-optimal minimax criterion is equivalent to the $D$-optimal criterion for selecting the optimal split-plot design without model misspecification.

\item[III.] When both $d=0$ and $\alpha=0$, equation (\ref{eq:ld}) reduces to $L_R({\cal D})=1/|\bX_1'\bX_1|$. Minimizing $L_R({\cal D})$ is equivalent to maximizing $|\bX_1'\bX_1|^{1/(p+1)}$ and hence the $D$-optimal minimax criterion is equivalent to the $D$-optimal criterion for selecting the optimal completely randomized design without model misspecification. 
\end{itemize}

\subsection{Scale invariance}
The $D$-optimal minimax criterion with the loss function as equation (\ref{eq:ld}) is scale invariant. Suppose that contrasts of the effects are rescaled by 
\[
\tilde\bH=\{\tilde\bH_1,\tilde\bH_2\}=\{\bH_1\bC_1,\bH_2\bC_2\},
\]
where $\bC_1=diag(1,c_1,\cdots,c_p)$, $\bC_2=diag(c_{p+1},\cdots,c_{N-1})$, and $c_1,\cdots,c_{N-1}$ are positive constants. Then 
$\tilde\bV_1=\bC_1\bV_1\bC_1$ and the fitted model with the rescaled effects can be written as
\begin{equation}\label{eq:m2}
\bY=\tilde\bX_1\boldsymbol\beta_1+\bZ\gamma+\epsilon,
\end{equation}
where $\tilde\bX_1=\bX_1\bC_1$. Let $L_{R(\bX_1)}({\cal D})$ and $L_{R(\tilde\bX_1)}({\cal D})$ represent the loss functions of $\cal D$ corresponding to model (\ref{eq:m1}) and model (\ref{eq:m2}), respectively. Then we obtain %(see Appendix~\ref{ap:sc})
\begin{equation}\label{eq:sc}
L_{R(\tilde\bX_1)}({\cal D})=L_{R(\bX_1)}({\cal D})/\prod_{i=1}^pc_i^2.
\end{equation}
If $\cal D$ is a $D$-optimal minimax design for model (\ref{eq:m1}), it minimizes $L_{R({\bX}_1)}({\cal D})$. Thus $\tilde{\bX}_1=\bX_1\bC_1$ minimizes $L_{R(\tilde{\bX}_1)}({\cal D})$ and is a $D$-optimal minimax design for model (\ref{eq:m2}). 
%It implies that the optimal split-plot designs selected by the $D$-optimal minimax criterion with different effect coding would be identical. 
Therefore, the $D$-optimal minimax criterion is scale invariant. 

\iffalse
\[
\begin{array}{rl}
l_A&=\max_{\frac{1}{N}\bo\beta_2'V_2\bo\beta_2\leq\alpha^2}tr(MSE)\\
&=tr((\bX_1'{\bf \Sigma}^{-1}{\bf X}_1)^{-1})(1+N\alpha^2\lambda_{max}((\bX_1'{\bf \Sigma}^{-1}{\bf X}_1)^{-1}\bX_1'{\bf \Sigma}^{-1}{\bf \Sigma}^{-1}{\bf X}_1-\bV_1^{-1})\\
\end{array}
\]
\fi

\section{Algorithm and update formulas}\label{se:au}
To construct and search for the robust split-plot design, we develop an efficient algorithm by combining the anneal algorithm and the point-exchange algorithm. 
The annealing algorithm has been shown effective for constructing the $D$-optimal designs or $D$-optimal minimax completely randomized designs (see Fang and Wines, 2002; Haines, 1987; Meyer and Nachtsheim, 1988; Zhou, 2001, 2008, 2011) and
the point-exchange algorithm is efficient for constructing optimal split-plot designs (see Goos and Vandebroek, 2001, 2003, 2004).
Another important issue for constructing the robust split-plot design is that the computing is intensive to calculate the inverse and determinant of the updated designs.  
To save the computational cost, we apply and modify the update formulas suggested by Arnouts and Goos (2010).

\subsection{Design construction algorithm}\label{se:al}
%To construct and search for the robust split-plot design, we combine the annealing algorithm and the coordinate-exchange algorithm. The annealing algorithm has been shown being effective for construct optimal 

%We combine two algorithms for constructing and finding the optimal $D$-optimal  xxx. The first is the annealing algorithm, which have been shown being effective for constructing optimal $D$-optimal or minimax robust designs, see Fang and Wines (2002), Haines (1987), Meyer and Nachtsheim (1998), and Zhou (2001, 2008, 2011).
%The second algorithm is coordinate exchange algorithm. This algorithm is efficient for constructing optimal split-plot designs which is candidate-set free, proposed to improve the point-exchange algorithm for large number of factors. 

Let $A$ be the candidate set of the whole plots, $E$ be the candidate set of the subplots, $T_0$ be the initial temperature, $a_b$ ($\leq b$) and $e_{n_i}$ ($\leq n_i$), $i=1,\cdots, b$, be the maximum numbers of whole plots and subplots that are allowed to change in a design to generate a new design, $N_T$ be the number of designs searched at each temperature, and $M_0$ be the total number of temperature changes.

\begin{itemize}
\item[] Step 1. Randomly generate an initial split-plot design ${\cal D}_0$ by selecting $b$ whole plots $\bw_i$, $i=1,\cdots,b$, from the candidate set of whole plots $A$ and selecting $n_i$ subplot $\bt_{ij}$, $j=1,\cdots,n_i$, for the $i$th whole plot from the candidate set of subplots $E$. Make sure that the design point $(\bw_i,\bt_{ij})$ for $i=1,\cdots,b$, $j=1,\cdots n_i$ are selected without replacement. Let $J$ be the number of temperature changes in the algorithm and set $J=1$ at beginning.

\item[] Step 2. Compute the loss function $L_R({\cal D}_0)$. %, where $\bX_1$ is the model expansion of $D_0$ with respect to the requirement set $R$ 
For each $i$, define a subset $E_i$ including all the points in $E$ that are not $\bt_{ij}$, $j=1,\cdots,n_i$.

\item[] Step 3. Implement point exchange for the whole plots. 
\begin{itemize}
\item[(a)] Randomly choose a number $a$ from set $\{1,\cdots,a_b\}$. Select $a$  whole plots $\bw_{i_l}$, $l=1,\cdots,a$, randomly and replace them by $a$ design points selected randomly from $A$ to obtain an updated design ${\cal D}^*$. 
\item[(b)] Compute the loss function $L_R({\cal D}^*)$. If $L_R({\cal D}^*)<L_R({\cal D}_0)$, then accept the updated design. Otherwise, accept the updated design ${\cal D}^*$ with probability $p_0=exp\{-[L_R({\cal D}^*)-L_R({\cal D}_0)]/T_0\}$. If the updated design is accepted, then let ${\cal D}_0={\cal D}^*$.     
\end{itemize}

\item[] Step 4. Conduct the point interchange for the whole plots. 
\begin{itemize}
\item[(a)] Swap design points between the $i$th whole plot and $l$th whole plot, where  $i\neq l$, to obtain an updated design ${\cal D}^*$. 
\item[(b)] Same as  Step 3 (b).
\end{itemize}

\item[] Step 5. Perform the point exchange for the subplots. 
\begin{itemize}
\item[(a)] For the $i$th whole plot, $i=1,\cdots,b$, randomly choose a number $e$ from set $\{1,\cdots,e_{n_i}\}$. Select $e$ subplot $\bt_{ij_l}$, $l=1,\cdots,e$, randomly and replace them by $e$ design points selected randomly from $E_i$ to obtain an updated design ${\cal D}^*$. 
\item[(b)] Same as Step 3 (b).
\end{itemize}

\item[] Step 6. Repeat Step 2 to Step 5 $N_T$ times and then go to Step 7.

\item[] Step 7. Reduce the temperature by a positive factor $f$ ($<1$), i.e., $T_0=fT_0$, and set $J=J+1$. If $J\leq M_0$, then go to Step 2. Otherwise, finish the process.

\end{itemize}

The final design ${\cal D}_0$ obtained from the algorithm can be consider as the $D$-optimal minimax split-plot design. %When $a_1=e_1=1$, Step 3 to Step 5 perform the point-exchange to improve the initial design. 

\subsection{Update formulas}
%Another important issue for constructing the robust split-plot design is that the computing is intensive to calculate the inverse and determinant of the updated designs.  
%To save the computational cost, we apply two update formula suggested by Arnouts and Goos (2010). 
If a matrix can be expressed in the form 
\begin{equation*}\label{eq:M}
\bM+\bQ\bD\bP,
\end{equation*}
then we can use the following formulas to calculate the determinant and the inverse of the matrix:
\begin{equation}\label{eq:up1}
|\bM+\bQ\bD\bP|=|\bM||\bD||\bD^{-1}+\bP\bM^{-1}\bQ|,
\end{equation} 
and 
\begin{equation}\label{eq:up2}
(\bM+\bQ\bD\bP)^{-1}=\bM^{-1}-\bM^{-1}\bQ(\bD^{-1}+\bP\bM^{-1}\bQ)^{-1}\bP\bM^{-1}.
\end{equation}
The second formula is called the Sherman-Morrison-Woodbury formula and the proofs of them were given by Harville (1997).

To apply the two formulas, let $\bM_1=\bX_1'\bo\Sigma^{-1}\bX_1$, $\bM_2=\bG_1'\bo\Sigma^{-1}\bG_1$, and $\bM_3= \bG_1'\bo\Sigma^{-2}\bG_1$, where $\bG_1=\bX_1\bV_1^{-\frac{1}{2}}$. Then equation (\ref{eq:ld}) can be written as
\begin{equation}\label{eq:ld1}
L_R({\cal D})=\frac{1+N\alpha^2\phi_R({\cal D})}{\pi_R({\cal D})},
\end{equation}
where $\phi_R({\cal D})=\lambda_{max}(\bM_2^{-1}\bM_3-\bM_2)$ and $\pi_R({\cal D})=|\bM_1|$.
%
%The inverse of the variance-covariance matrix can be expressed by $\bo\Sigma^{-1}=\sigma_\epsilon^{-2}\bI_n-\sigma_\epsilon^{-2}d\bZ(\bI_b+d\bZ'\bZ)^{-1}\bZ'$ and $\bo\Sigma^{-2}=\bo\Sigma^{-1}\bo\Sigma^{-1}=\sigma_\epsilon^{-2}\bI_n-2\sigma_\epsilon^{-4}d\bZ(\bI_b+d\bZ'\bZ)^{-1}\bZ'+\sigma_\epsilon^{-4}d^2\bZ(\bI_b+d\bZ'\bZ)^{-1}\bZ'\bZ(\bI_b+d\bZ'\bZ)^{-1}$.
%
%If runs in the split-plot design is ordered by the whole plot, we can write the above equations by $\bo\Sigma^{-1}=diag(\bo\Sigma^{-1}_1,\cdots,\bo\Sigma^{-1}_b)$ and $\bo\Sigma^{-2}=diag(\bo\Sigma^{-2}_1,\cdots,\bo\Sigma^{-2}_b)$ where $\bo\Sigma^{-1}_i=\sigma_\epsilon^{-2}\bI_{n_i}-\sigma_\epsilon^{-2}\frac{d}{1+dn_i}{\bf 1}_{n_i}{\bf 1}_{n_i}'$ and 
%$\bo\Sigma^{-2}_i=\sigma_\epsilon^{-4}\bI_{n_i}+\sigma_\epsilon^{-4}\frac{d}{1+dn_i}(\frac{dn_i}{1+dn_i}-2){\bf 1}_{n_i}{\bf 1}_{n_i}'$
% 
Let $\bX_{1i}$ and $\bG_{1i}$ be the submatrices of $\bX_1$ and $\bX_1\bV_1^{-\frac{1}{2}}$, respectively, corresponding to the $i$th whole plot. Let $\bff(\bw_i,\bt_{ij})$ and $\bg(\bw_i,\bt_{ij})$ be the rows of $\bX_1$ and $\bX_1\bV_1^{-\frac{1}{2}}$, respectively, corresponding to the design point of the $j$th subplot in the $i$th whole plot. 
%
%Suppose that the point exchange or interchange process of the algorithm updates $M_1$, $M_2$, $M_3$ to $M_1^*$, $M_2^*$, $M_3^*$. 
%Since $M_1^*$, $M_2^*$, and $M_3^*$ can be expressed in the form of equation (\ref{eq:M}), the update 
If the point exchange or interchange process of the algorithm updates $\bM_1$, $\bM_2$, $\bM_3$ to $\bM_1^*$, $\bM_2^*$, $\bM_3^*$, and %the number of exchanged points is not too large, i.e.,
$a$ ane $e$ are small, then formulas (\ref{eq:up1}) and (\ref{eq:up2}) are efficient for calculating the invariances and determinants of $\bM_1^*$, $\bM_2^*$, $\bM_3^*$ for the updated design. 
We summarize the results of the update formulas for Step 3 to Step 5 in Table~\ref{tb:up}. %The notations are described as follows and the details are given in Appendix~\ref{ap:up}.

\begin{table}[h]
\begin{center}
\caption{Update formulas for Step 3 to Step 5 of the algorithm}\label{tb:up}
\begin{tabular}{rl}
\hline
(a)&Update formulas\\
&$\begin{array}{l@{=}l}
|\bM_1^{*}|&|\bM_1||\bD_1||\bD_1^{-1}+\bP_1\bM_1^{-1}\bP_1'|\\
\bM_2^{*}&\bM_2+\bP_2\bD_1\bP_2'\\
\bM_2^{*-1}&\bM_2^{-1}-\bM_2^{-1}\bP_2'(\bD_1^{-1}+\bP_2\bM_2^{-1}\bP_2')^{-1}\bP_2\bM_2^{-1}\\
\bM_3^{*}&\bM_3+\bP_2\bD_2\bP_2'\\
\end{array}$\\
(b)&Step 3. Exchange for whole plots\\
&$\begin{array}{l@{=}l}
\bP_1&(\bX_{1i_1}',\cdots,\bX_{1i_a}',\bX_{1i_1}^{*'},\cdots,\bX_{1i_a}^{*'},\bX_{1i_1}'{\bf 1}_{n_{i_1}},\cdots,\bX_{1i_a}'{\bf 1}_{n_{i_a}},\bX_{1i_1}^{*'}{\bf 1}_{n_{i_1}},\cdots,\bX_{1i_a}^{*'}{\bf 1}_{n_{i_a}})'\\
\bD_1&\sigma_\epsilon^{-2}diag(-\bI_{n_{i_1}},\cdots,-\bI_{n_{i_a}},\bI_{n_{i_1}},\cdots,\bI_{n_{i_a}},\frac{d}{1+dn_{i_1}},\cdots,\frac{d}{1+dn_{i_a}},-\frac{d}{1+dn_{i_1}},\cdots,-\frac{d}{1+dn_{i_a}})\\
\bP_2&(\bG_{1i_1}',\cdots,\bG_{1i_a}',\bG_{1i_1}^{*'},\cdots,\bG_{1i_a}^{*'},\bG_{1i_1}'{\bf 1}_{n_{i_1}},\cdots,\bG_{1i_a}'{\bf 1}_{n_{i_a}},\bG_{1i_1}^{*'}{\bf 1}_{n_{i_1}},\cdots,\bG_{1i_a}^{*'}{\bf 1}_{n_{i_a}})'\\
\bD_2&\sigma_\epsilon^{-4}diag(-\bI_{n_{i_1}},\cdots,-\bI_{n_{i_a}},\bI_{n_{i_1}},\cdots,\bI_{n_{i_a}},\frac{2d+d^2n_{i_1}}{(1+dn_{i_1})^2},\cdots,\frac{2d+d^2n_{i_a}}{(1+dn_{i_a})^2},-\frac{2d+d^2n_{i_1}}{(1+dn_{i_1})^2},\cdots,-\frac{2d+d^2n_{i_a}}{(1+dn_{i_a})^2})\\
\end{array}$\\
(c)&Step 4. Interchange for whole plots\\
&$\begin{array}{l@{=}l}
\bP_1&(\bX_{1i}'{\bf 1}_{n_i},\bX_{1l}'{\bf 1}_{n_l},\bX_{1i}^{*'}{\bf 1}_{n_i},\bX_{1l}^{*'}{\bf 1}_{n_l})'\\ 
\bD_1&\sigma_\epsilon^{-2}diag(\frac{d}{1+dn_i},\frac{d}{1+dn_l},-\frac{d}{1+dn_i},-\frac{d}{1+dn_l})\\ 
\bP_2&(\bG_{1i}'{\bf 1}_{n_i},\bG_{1l}'{\bf 1}_{n_l},\bG_{1i}^{*'}{\bf 1}_{n_i},\bG_{1l}^{*'}{\bf 1}_{n_l})'\\
\bD_2&\sigma_\epsilon^{-4}diag(\frac{2d+d^2n_i}{(1+dn_i)^2},\frac{2d+d^2n_l}{(1+dn_l)^2},-\frac{2d+d^2n_i}{(1+dn_i)^2},-\frac{2d+d^2n_l}{(1+dn_l)^2})\\
\end{array}$\\
(d)&Step 5. Exchange for subplots\\
&$\begin{array}{l@{=}l}
\bP_1&(\bff(\bw_i,\bt_{ij_1}),\cdots,\bff(\bw_i,\bt_{ij_e}),\bff(\bw_i,\bt_{ij_1}^*),\cdots,\bff(\bw_i,\bt_{ij_1}^*),\bX_{1i}'{\bf 1}_{n_i},\bX_{1i}^{*'}{\bf 1}_{n_i})'\\ 
\bD_1&\sigma_\epsilon^{-2}diag(-1,\cdots,-1,1,\cdots,1,\frac{d}{1+dn_i},-\frac{d}{1+dn_i})\\ 
\bP_2&(\bg(\bw_i,\bt_{ij_1}),\cdots,\bg(\bw_i,\bt_{ij_e}),\bg(\bw_i,\bt_{ij_1}^*),\cdots,\bg(\bw_i,\bt_{ij_e}^*),\bG_{1i}'{\bf 1}_{n_i},\bG_{1i}'{\bf 1}_{n_i})'\\ 
\bD_2&\sigma_\epsilon^{-4}diag(-1,\cdots,-1,1,\cdots,1,\frac{2d+d^2n_{i}}{(1+dn_{i})^2},-\frac{2d+d^2n_{i}}{(1+dn_{i})^2})\\
\end{array}$\\
\hline
\end{tabular}
\end{center}
\end{table}

In Step 3, assume that the design point of whole-plot factors $\bw_{i_l}$ is substituted by $\bw_{i_l}^*$, $l=1,\cdots,a$. This point exchange for whole-plot factors updates $\bX_1$, $\bG_1$, $\bM_1$, $\bM_2$, and $\bM_3$ to $\bX_1^{*}$, $\bG_1^{*}$, $\bM_1^{*}$, $\bM_2^{*}$, and $\bM_3^{*}$. Let $\bX_{1i_l}^{*}$ and $\bG_{1i_l}^{*}$ be the submatrices of $\bX_1^{*}$ and $\bG_1^{*}$, respectively, corresponding to the $i_l$th whole plot, $l=1,\cdots,a$.
Then equation (\ref{eq:ld1}) for the updated design in Step 3 can be calculated by replacing $|\bM_1|$, $\bM_2$, $\bM_2^{-1}$, and $\bM_3$ by $|\bM_1^{*}|$, $\bM_2^*$, $\bM_2^{*-1}$, and $\bM_3^{*}$, respectively, listed in Table~\ref{tb:up}~(a) where $\bP_1$, $\bD_1$, $\bP_2$, and $\bD_2$ are given in Table~\ref{tb:up}~(b).

In Step 4, assume that the design points of whole-plot factors $\bw_{i}$ and $\bw_{l}$ are interchanged. This point interchange for whole-plot factors updates $\bX_1$, $\bG_1$, $\bM_1$, $\bM_2$, and $\bM_3$ to $\bX_1^{*}$, $\bG_1^{*}$, $\bM_1^{*}$, $\bM_2^{*}$, and $\bM_3^{*}$.
Let $\bX_{1i}^{}$ and $\bX_{1l}^{}$ ($\bG_{1i}^{}$ and $\bG_{1l}^{}$) be the submatrices of $\bX_1$ ($\bG_1$) corresponding to the $i$th and $l$th whole plots, respectively, and $\bX_{1i}^{*}$ and $\bX_{1l}^{*}$ ($\bG_{1i}^{*}$ and $\bG_{1l}^{*}$) be the submatrices of $\bX_1^{*}$ ($\bG_1^{*}$) corresponding to the $i$th and $l$th whole plots, respectively.
Then equation (\ref{eq:ld1}) for the updated design in Step 4 can be calculated by replacing $|\bM_1|$, $\bM_2$, $\bM_2^{-1}$, and $\bM_3$ by $|\bM_1^{*}|$, $\bM_2^*$, $\bM_2^{*-1}$, and $\bM_3^{*}$, respectively, listed in Table~\ref{tb:up}~(a) where $\bP_1$, $\bD_1$, $\bP_2$, and $\bD_2$ are given in Table~\ref{tb:up}~(c).

In Step 5, assume that the design point of subplot factors $\bt_{ij_l}$ is substituted by $\bt_{ij_l}^*$, $l=1,\cdots,e$. This point exchange for subplot factors updates $\bX_1$, $\bG_1$, $\bM_1$, $\bM_2$, and $\bM_3$ to $\bX_1^{*}$, $\bG_1^{*}$, $\bM_1^{*}$, $\bM_2^{*}$, and $\bM_3^{*}$. Let $\bX_{1i_l}^{*}$ and $\bG_{1i_l}^{*}$ be the submatrices of $\bX_1^{*}$ and $\bG_1^{*}$, respectively, corresponding to the $i$th whole plot and $\bff(\bw_i,\bt_{ij_l}^*)$ and $\bg(\bw_i,\bt_{ij_l}^*)$ be the rows of $\bX_1^{*}$ and $\bG_1^{*}$, respectively, corresponding to the $j_l$th design point in the $i$th whole plot, $l=1,\cdots,e$. Then equation (\ref{eq:ld1}) for the updated design in Step 5 can be calculated by replacing $|\bM_1|$, $\bM_2$, $\bM_2^{-1}$, and $\bM_3$ by $|\bM_1^{*}|$, $\bM_2^*$, $\bM_2^{*-1}$, and $\bM_3^{*}$, respectively, listed in Table~\ref{tb:up}~(a) where $\bP_1$, $\bD_1$, $\bP_2$, and $\bD_2$ are given in Table~\ref{tb:up}~(d).

\section{Examples}\label{se:ex}
We apply the construction algorithm developed in Section~\ref{se:al} and use the general form of the loss function for $D$-optimal minimax criterion introduced in Section~\ref{se:ld} to obtain the robust split-plot designs for model misspecification. Two examples are given to demonstrate our methods. The first example is for two-level split-plot designs and the second example is for mixed-level split-plot designs. 

\begin{table}[h]
\caption{$D$-optimal split-plot design (${\cal D}_1$ with $\alpha=0$) and $D$-optimal minimax split-plot design (${\cal D}_2$ with $\alpha=1$) for $\sigma_\epsilon^2=\sigma_\gamma^2=1$.}\label{tb:ex1}
\begin{center}
\begin{tabular}{cc}											
$\begin{array}{c|rrrrr}											
\multicolumn{6}{c}{\mbox{Design }{\cal D}_1}\\											
\hline											
WP	&	F_1	&	F_2	&	F_3	&	F_4	&	F_5	\\
\hline											
1	&	1	&	-1	&	1	&	1	&	1	\\
	&	1	&	-1	&	-1	&	-1	&	-1	\\
	&	1	&	-1	&	-1	&	-1	&	1	\\
	&	1	&	-1	&	1	&	1	&	-1	\\
\hline											
2	&	-1	&	1	&	1	&	1	&	-1	\\
	&	-1	&	1	&	1	&	-1	&	-1	\\
	&	-1	&	1	&	-1	&	1	&	1	\\
	&	-1	&	1	&	-1	&	-1	&	1	\\
\hline											
3	&	-1	&	-1	&	-1	&	-1	&	-1	\\
	&	-1	&	-1	&	1	&	-1	&	1	\\
	&	-1	&	-1	&	-1	&	1	&	-1	\\
	&	-1	&	-1	&	1	&	1	&	1	\\
\hline											
4	&	1	&	1	&	1	&	-1	&	-1	\\
	&	1	&	1	&	-1	&	1	&	1	\\
	&	1	&	1	&	-1	&	1	&	-1	\\
\hline											
\multicolumn{6}{l}{\phi_R({\cal D}_1)=.6733}\\											
\multicolumn{6}{l}{\pi_R({\cal D}_1)^{1/(1+p)}=6.7468}\\											
\multicolumn{6}{l}{L_R({\cal D}_1)^{1/(1+p)}=.2188}\\											
\end{array}$											
&											
$\begin{array}{c|rrrrr}											
\multicolumn{6}{c}{\mbox{Design }{\cal D}_2}\\											
\hline											
WP	&	F_1	&	F_2	&	F_3	&	F_4	&	F_5	\\
\hline											
1	&	1	&	1	&	-1	&	1	&	1	\\
	&	1	&	1	&	1	&	-1	&	-1	\\
	&	1	&	1	&	1	&	1	&	-1	\\
	&	1	&	1	&	-1	&	-1	&	1	\\
\hline											
2	&	-1	&	-1	&	-1	&	-1	&	1	\\
	&	-1	&	-1	&	1	&	-1	&	-1	\\
	&	-1	&	-1	&	1	&	1	&	1	\\
	&	-1	&	-1	&	-1	&	1	&	-1	\\
\hline											
3	&	1	&	-1	&	-1	&	-1	&	-1	\\
	&	1	&	-1	&	1	&	-1	&	1	\\
	&	1	&	-1	&	-1	&	1	&	-1	\\
	&	1	&	-1	&	1	&	1	&	1	\\
\hline											
4	&	-1	&	1	&	-1	&	-1	&	1	\\
	&	-1	&	1	&	1	&	1	&	-1	\\
	&	-1	&	1	&	-1	&	1	&	-1	\\
\hline											
\multicolumn{6}{l}{\phi_R({\cal D}_2)=.6323}\\											
\multicolumn{6}{l}{\pi_R({\cal D}_2)^{1/(1+p)}=6.7339}\\											
\multicolumn{6}{l}{L_R({\cal D}_2)^{1/(1+p)}=.2176}\\											
\end{array}$											
\end{tabular}																						
\end{center}
\end{table}

\iffalse
\begin{figure}%[h]
\centering
\begin{minipage}{.4\textwidth}
\includegraphics[width=\textwidth]{a=0,d=1-1.pdf}
\caption{Design ${\cal D}_1$}
\end{minipage}
\begin{minipage}{.4\textwidth}
\includegraphics[width=\textwidth]{a=1,d=1-1.pdf}
\caption{Design ${\cal D}_2$}
\end{minipage}
\end{figure}
\fi

\begin{figure}
\begin{subfigure}{.5\textwidth}
  \centering
  \includegraphics[width=.8\linewidth]{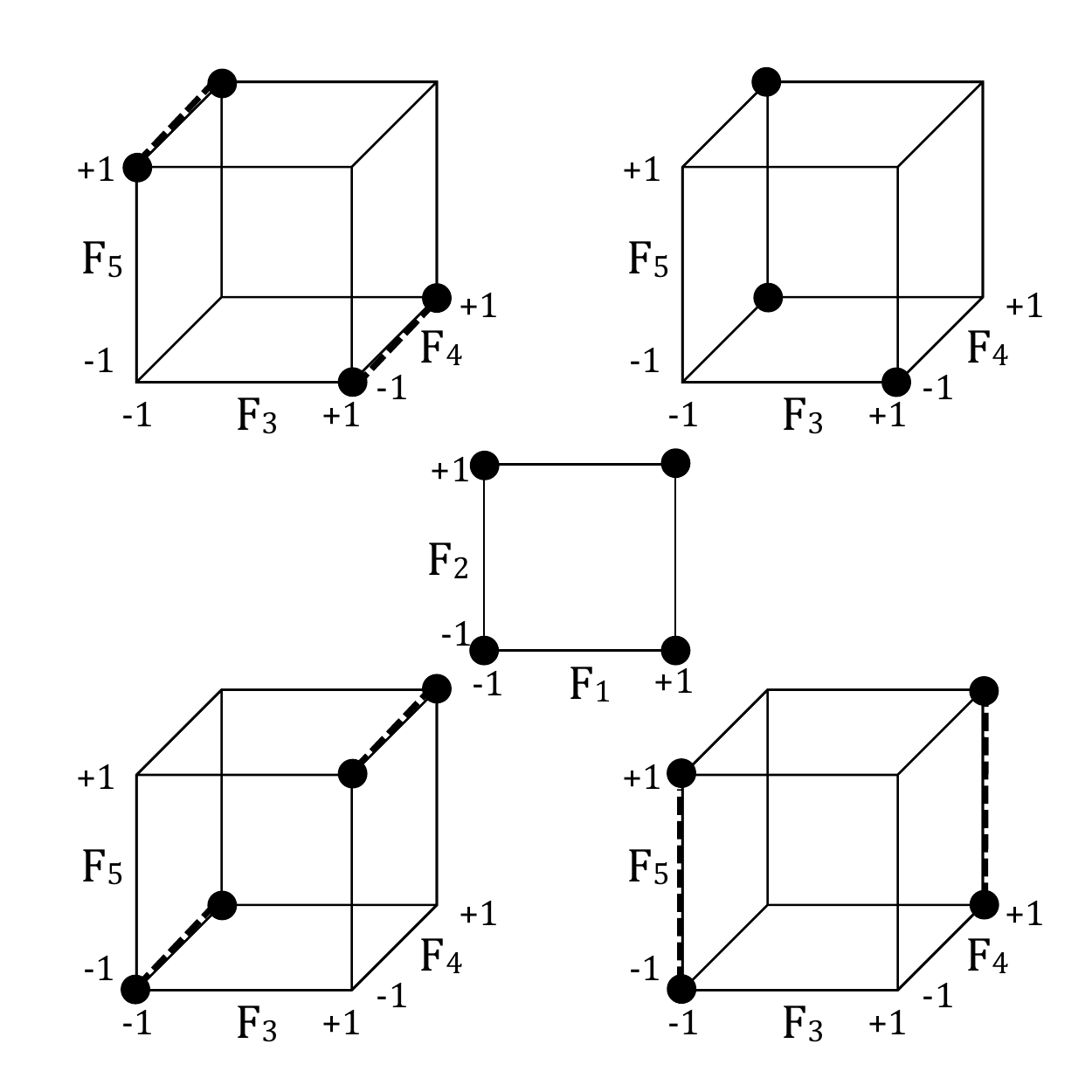}
  \caption{Design ${\cal D}_1$}
  \label{fig:sfig1}
\end{subfigure}%
\begin{subfigure}{.5\textwidth}
  \centering
  \includegraphics[width=.8\linewidth]{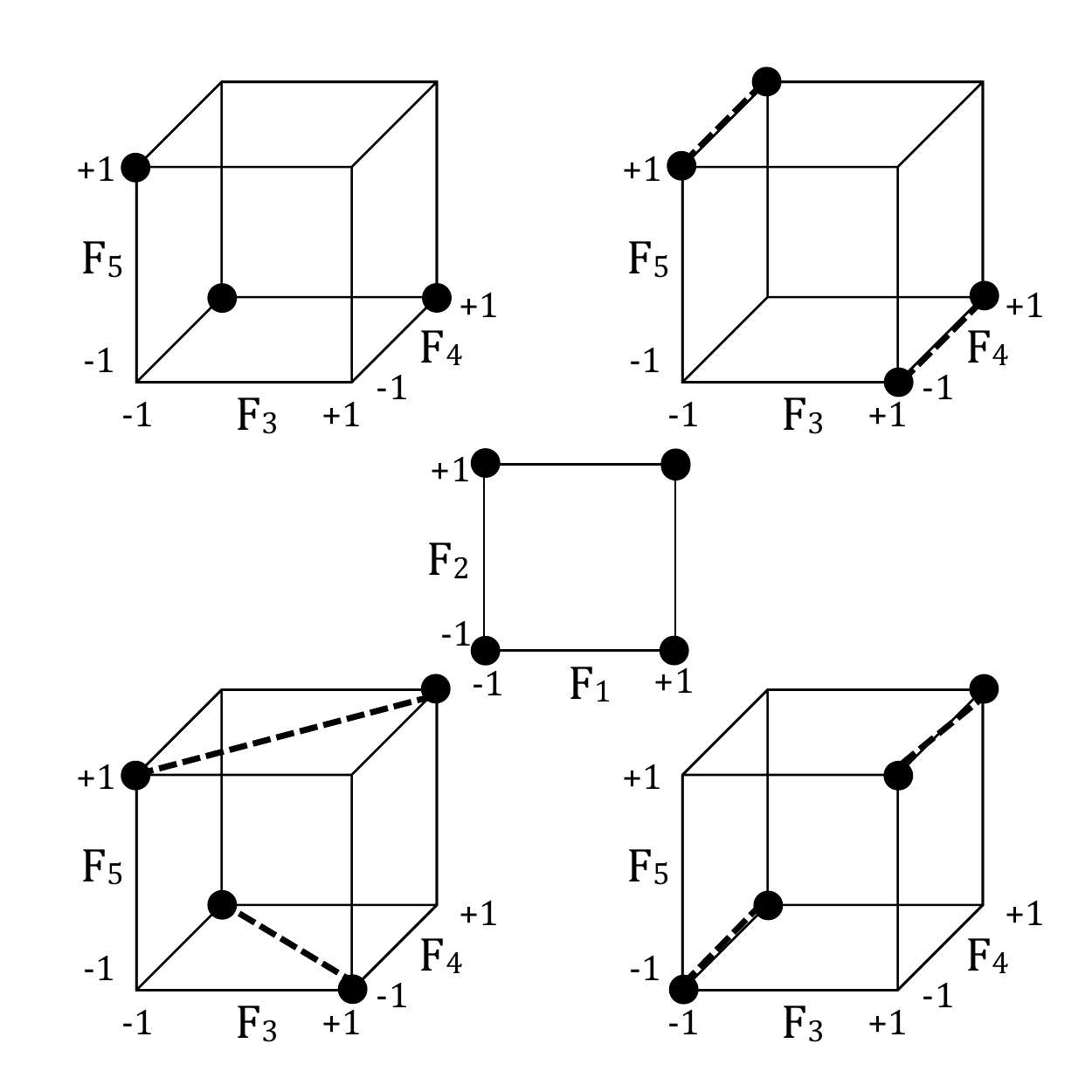}
  \caption{Design ${\cal D}_2$}
  \label{fig:sfig2}
\end{subfigure}
\caption{Structures of ${\cal D}_1$ and ${\cal D}_2$}
\label{fi:ex1}
\end{figure}

\textbf{Example 1.} %Assume that an experiment are performed for analyzing four potentially important factors. If two of the four factors are difficult to change level, a split-plot design is recommended. 
Consider to construct a split-plot design with fifteen runs and five two-level factors $F_1,\cdots,F_5$, where the two levels of factors are coded as $(-1,+1)$. 
Suppose that the first two factors $F_1$ and $F_2$ are hard-to-change factors arranged into four whole plots ($b=4$) and the last three factors $F_3,$ $F_4$, $F_5$ are easy-to-change factors where the numbers of subplots in the $i$th whole plot are $n_i=4$ for $i=1,2,3$ and $n_i=3$ for $i=4$.
If we want to investigate all the main effects, the interaction of $F_1$ and $F_2$, and the interaction of $F_1$ and $F_3$, then the requirement set is $R=\{x_1,x_2,x_3,x_4,x_5,x_1x_2,x_1x_3\}$, where $x_i$ is the main effect of $F_i$, $i=1,\cdots,5$. 
The number of effects in the requirement set $R$ is $p=7$ and $N$ in this case is $2^5=32$.
Let the candidate set of whole plots $A$ be the $2^2$ full factorial design and the candidate set of subplots $E$ be the $2^3$ full factorial design.  
We apply the algorithm by setting $T_0=.001$, $M_0=50$, $N_T=100$, $a_b=3$, $e_{n_i}=3$, and $f=.8$ to search for the $D$-optimal minimax split-plot design ($\alpha=1$) and the $D$-optimal split-plot design ($\alpha=0$) with $\sigma_\epsilon=1$ and $\sigma_\gamma=1$.
The update formulas given in Table~\ref{tb:up} are used for increasing the computing speed for obtaining the loss function of the updated designs, $L_R({\cal D}^*)$.
Table~\ref{tb:ex1} lists two optimal split-plot designs. 
The $(1+p)$th root of the determinant of the information matrix of ${\cal D}_1$ with respect to the requirement set $R$ is $\pi_R({\cal D}_1)^{1/(1+p)}=6.7468$. Since $\pi_R({\cal D}_1)^{1/(1+p)}$ is minimum among all of the designs we construct, ${\cal D}_1$ is the $D$-optimal split-plot design. 
However, if the fitted model is misspecified and there exist small departures from the underlying true model with $\alpha=1$, 
then the $(1+p)$th root of the loss function of ${\cal D}_1$ is $L_R({\cal D}_1)^{1/(1+p)}=.2188$, which is greater than $L_R({\cal D}_2)^{1/(1+p)}=.2176$.
It implies that design ${\cal D}_1$ has higher bias of the estimation than design ${\cal D}_2$. 
%It implies that design ${\cal D}_1$ is not the best robust design for the model misspecification.  
%
Since $L_R({\cal D}_2)^{1/(1+p)}$ is minimum among all of the designs we construct, ${\cal D}_2$ is the $D$-optimal minimax split-plot design. Therefore, when  $\alpha=1$, design ${\cal D}_2$ is the optimally robust split-plot design for model misspecification. 
We look into the allocations of the design points in ${\cal D}_1$ and ${\cal D}_2$ and show their structures in Figure~\ref{fi:ex1}.
We find that, in each whole plot with $n_i=4$ in ${\cal D}_1$, the connection of two points of the subplots can be parallel to the connection of the other two points. However, for the whole plot $(-1,-1)$ in ${\cal D}_2$, it is impossible to connect two points which is parallel to the connection of the other two points. The different structures between the two designs result in less bias of the estimation for design ${\cal D}_2$ when the model is misspecified with $\alpha=1$.

%while in one the whole plots with $n_i=4$ for $D_2$, its four subplots are allocated in to different planes. 
%This might be the reason that the two design have 

%
%In fact, the minimum eigenvalues for the two designs are $\phi_R(D_1)=.6733$ and $\phi_R(D_2)=.6323$, and the determinants are $\pi_R(D_1)=4,293,967$ and $\pi_R(D_2)=4,227,858$. We can obtain $\alpha_0=(\frac{\pi_R(D_2)-\pi_R(D_1)}{N(\pi_R(D_1)\phi_R(D_2)-\pi_R(D_2)\phi_R(D_1))})^{1/2}=.1266$.
%
%If $\frac{1}{N}\bo\beta_2'\bV_2\bo\beta_2\leq\alpha_0^2$

\iffalse
while those of design $D_2$ are $\phi_R(D_2)=.6323$ and $\pi_R(D_2)=4,227,858$.
Design $D_1$ has $\phi_R(D_1)=.6323$ and $\pi_R(D_1)=4,227,858$. Design $D_2$ has $\phi_R(D_2)=.6733$ and $\pi_R(D_2)=4,294,967$,
%
The $\phi_2$ of $D_2$ is maximum and hence is $D$-optimal split-plot design. However, the $\phi_1$ of $D_1$ is less than $\phi_1$ of $D_2$. In there exist a differ with $\alpha\leq .1266$, $D_2$ is also the $D$-optimal minimax split-plot design. When $\alpha> .1266$, $D_1$ is also the $D$-optimal minimax split-plot design. 
\fi

%\begin{figure}
%%\begin{\center}
%\center
%\includegraphics[width=4in]{a=1,d=1.pdf}
%\includegraphics[width=4in]{a=0,d=1.pdf}
%%\end{center}
%\caption{The face centered CCD and the projections of DS.17 and $OA(18,3^7)$ on $k=3$.}
%\label{fi:pro}
%\end{figure}

\begin{table}
\caption{Designs ${\cal D}_3$ and ${\cal D}_4$ in Example 2.} 
\begin{center}
\begin{tabular}{cc}							
$\begin{array}{c|rrr}							
\multicolumn{4}{c}{\mbox{Design }{\cal D}_3}\\							
\hline							
WP	&	F_1	&	F_2	&	F_3	\\
\hline							
1	&	-1	&	0	&	0	\\
	&	-1	&	1	&	1	\\
\hline							
2	&	1	&	0	&	2	\\
	&	1	&	0	&	1	\\
\hline							
3	&	-1	&	2	&	2	\\
	&	-1	&	2	&	0	\\
	&	-1	&	0	&	1	\\
\hline							
4	&	1	&	2	&	1	\\
	&	1	&	1	&	2	\\
	&	1	&	1	&	0	\\
\hline							
\multicolumn{4}{l}{\phi_R({\cal D}_3)=.9074}\\							
\multicolumn{4}{l}{\pi_R({\cal D}_3)^{1/(1+p)}=4.5472}\\							
\multicolumn{4}{l}{L_R({\cal D}_3)^{1/(1+p)}=.2925}\\							
\end{array}$							
&							
$\begin{array}{c|rrr}							
\multicolumn{4}{c}{\mbox{Design }{\cal D}_4}\\							
\hline							
WP	&	F_1	&	F_2	&	F_3	\\
\hline							
1	&	-1	&	2	&	2	\\
	&	-1	&	1	&	1	\\
\hline							
2	&	1	&	2	&	0	\\
	&	1	&	0	&	2	\\
\hline							
3	&	-1	&	1	&	0	\\
	&	-1	&	0	&	2	\\
	&	-1	&	2	&	1	\\
\hline							
4	&	1	&	2	&	1	\\
	&	1	&	0	&	0	\\
	&	1	&	1	&	2	\\
\hline							
\multicolumn{4}{l}{\phi_R({\cal D}_4)=.6667}\\							
\multicolumn{4}{l}{\pi_R({\cal D}_4)^{1/(1+p)}=4.5472}\\							
\multicolumn{4}{l}{L_R({\cal D}_4)^{1/(1+p)}=.2842}\\							
\end{array}$							
\end{tabular}										
\end{center}
\end{table}

\textbf{Example 2.}
Consider an experiment with factors $F_1$, $F_2$, and $F_3$, where $F_1$ has two levels and $F_2$ and $F_3$ have three levels. Assume that $F_1$ is a hard-to-change factor arranged into four whole plots. The other two factors $F_2$ and $F_3$ are easy to change and the numbers of subplots in the $i$th whole plot are $n_i=2$ for $i=1,2$ and $n_i=3$ for $i=3,4$. 
If we are interested in estimating all the main effects, the interaction between $F_1$ and $F_2$, and the interaction between $F_1$ and $F_3$, then the requirement set is $R=\{x_1,x_{2L},x_{2Q},x_{3L},x_{3Q},x_1x_{2L},x_1x_{2Q},x_1x_{3L},x_1x_{3Q}\}$, where $x_1$ is the main effect of $F_1$ and $x_{iL}$ and $x_{iQ}$ are the linear and quadratic components of the main effect of factor $F_i$, $i=2,3$. 
The two levels of $F_1$ are coded as $(-1,+1)$ and the three levels $(0,1,2)$ of $F_i$ are coded as $\sqrt \frac{3}{2}(-1,0,+1)$ for $x_{iL}$ and $\sqrt\frac{1}{2}(+1,-2,+1)$ for $x_{iQ}$, $i=2,3$. 
The candidate set of the whole plots $A$ is the $2^1$ full factorial design and the candidate set of the subplots $E$ is the $3^2$ full factorial design.    
We apply the algorithm by setting $T_0=.001$, $M_0=50$, $N_T=100$, $a_b=3$, $e_{n_i}=2$, and $f=.8$ to construct the $D$-optimal minimax split-plot design with $\sigma_\epsilon^2=1$, $\sigma_\gamma^2=1$, and $\alpha=1$.
Table 2 lists two designs ${\cal D}_3$ and ${\cal D}_4$. 
Both ${\cal D}_3$ and ${\cal D}_4$ have the same $(1+p)$th root of the determinants of the information matrices, $\pi_R({\cal D}_3)^{1/(1+p)}=\pi_R({\cal D}_4)^{1/(1+p)}=4.5472$. Since $\pi_R({\cal D}_3)^{1/(1+p)}$ and $\pi_R({\cal D}_4)^{1/(1+p)}$ are maximum among all of the designs, both ${\cal D}_3$ and ${\cal D}_4$ are $D$-optimal split-plot designs. 
However, if model is misspecified with $\alpha=1$, then design ${\cal D}_3$ has $\phi_R({\cal D}_3)=0.9074$ and ${\cal D}_4$ has $\phi_R({\cal D}_4)=0.6667$. The value of $\phi_R({\cal D}_4)$ is minimum among all of the designs. Therefore, ${\cal D}_4$ is the $D$-optimal minimax split-plot design.
Since $\pi_R({\cal D}_4)^{1/(1+p)}$ is maximum and $\phi_R({\cal D}_4)$ is minimum, the value of $L_R({\cal D}_4)$ does not depend on $\alpha$ and is minimum among all of the designs. 
Therefor, ${\cal D}_4$ is the optimally robust split-plot design with or without model misspecification. 
This example shows that the $D$-optimal minimax split-plot design could be also the $D$-optimal split-plot design.

\iffalse
to construct a mixed-level split-plot design with one two-level whole-plot factor coded by $(-1,+1)$ for levels (0,1) and two three-level subplot factors where the linear effects and the quadratic effects are coded as $(-1,0,+1)$ and $(+1,2,+1)$, respectively, for levels $(0,1,2)$. This design has for whole plots and 
Assume that the requirement set is $R=\{x_1,x_{2L},x_{2Q},x_{3L},x_{3Q},x_1x_{2L},x_1x_{2Q}\}$, where $x_1$ is the main effect of $F_3$ and $x_{iL}$ and $x_{iQ}$ are the linear and quadratic components of the main effect of factor $F_i$, $i=2,3$. 

The $D$-optimal minimax designs with $n=10$ and $(n_1,n_2,n_3,n_4)=(2,3,2,3)$ are constructed. Table 2 represents design $D_1$ and $D_2$, where $D_1$ is a $D$-optimal design but 
%Consider to construct a three-level and 30-run split-plot design for two whole-plot factors and three subplot factors with 10 whole plots and three subplot in each whole plot, that is, $n=30$, $m_w=2$, $m_s=3$, $b=10$, $n_i=3$, $i=1,\cdots,10$.
%Assume that a second-order model is of the interest. 
%The requirement set includes all the main effects and interaction effects, that is,   
%$R=\{x_{1L},\cdots,x_{5L},x_{1Q},\cdots,x_{5Q},x_{1L}x_{2L},\cdots,x_{4L}x_{5L}\}$ with $p=30$ effects. The linear components $x_{iL}$ and the quadratic components $x_{iQ}$ of factor $F_i$ can be coded as $(-1,0,+1)$ and $(+1,-2,+1)$, respectively, for levels $(0,1,2)$, $i=1,\cdots,5$. 
%We run the algorithm several times with different initial designs and choose the 
\fi

\section{Conclusions and remarks}\label{se:co}
Many approaches for constructing optimal split-plot designs could be found in literature. However, the optimal designs obtained by these methods might be unrobust for model misspecification. 
If there exist significant effects that are not included in the model, then the estimation of effects could be highly biased. 
In this paper, we take the model misspecification into account.
We extend the application of the $D$-optimal minimax criterion to the split-plot design and provide a general form of the loss function for the criterion.
This general form of the loss function can be used for finding the optimal design for split-plot experiments and complete randomized experiments with or without model misspecification.
By combining the anneal algorithm and the point-exchange algorithm, we develop a new construction algorithm to efficiently obtain the robust split-plot design for model misspecification.

There exist two articles that are related to our work. 
The first article by Smucker et al. (2012) provides a method to obtain the model-robust designs for split-plot experiments. The authors argued that many methods in literature rely on the a priori assumption that the form of the regression function is known. They relaxed this assumption by allowing a set of model forms to be specified. This method uses a scaled product of $D$-optimal criterion to produce designs that account for all models in the set. 
% 
%Although both of their method and our method are for finding robust split-plot designs, these two method are in very different concepts. 
%The concept of this method is different from the method we propose. 
%
This method is innovative and the optimal split-plot design constructed by it is robust if the specified set of model forms includes the true model. 
However, in practice the underlying true model is usually complicated and unknown. It is not easy to specify a set that including the unknown true model.
In this paper, we relax this constraint for the specified set. 
Our method dose not rely on the knowledge of the true model and allows the fitted model differing from the unknown true model.
By minimizing the loss function, the optimal split-plot design obtained by our method can simultaneously control the variances/covariances and the bias of the estimation and hence is robust for model misspecification.       
%
%this is usually the case, then the set of possible models may 
%how to determine the set of models is 
%only one model would be choose to fit the data. If the set of possible models does not include the u of models, then the optimal design selected by this method may not be robust  

%minimizing the overall determinant of the information matrices           

Another article by Mann et al. (2014) is close to our work. Both of their method and our method use the $D$-optimal minimax criterion but with different model setting. The method proposed by Mann et al. (2014) assumed that the block effects are fixed while our method assumes that the block effects are random. 
The former is usually used for finding the robust block design with model misspecification and the later is used for obtaining the robust split-plot design. 

In summary, a good split-plot design should be able to control both the variances/covariances and the bias of the estimation. The method we propose can achieve this goal and construct the robust split-plot design for model misspecification.

\end{document}